\documentclass{elsarticle}\journal{Computational Materials Science}\usepackage{hyperref}

\usepackage{amsmath,amsthm,amssymb}
\usepackage{bm}
\usepackage{graphicx}

\newcommand{\mA}{{\mathsf A}}

\newcommand{\bbR}{{\mathbb R}}
\newcommand{\bbZ}{{\mathbb Z}}

\newcommand{\eps}{{\epsilon}}

\newcommand{\cut}{{\rm cut}}
\newcommand{\qm}{{\rm qm}}

\begin{document}

\begin{frontmatter}

\author{A. Shapeev}
\address{Skolkovo Institute of Science and Technology,
Skolkovo Innovation Center, Nobel Str.\ 3,
Moscow 143026, Russia}


\title{Accurate representation of formation energies of crystalline alloys with many components}

\date{\today}

\begin{abstract}
In this paper I propose a new model for representing the formation energies of multicomponent crystalline alloys as a function of atom types.
In the cases when displacements of atoms from their equilibrium positions are not large, the proposed method has a similar accuracy as the state-of-the-art cluster expansion method, and a better accuracy when the fitting dataset size is small.
The proposed model has only two tunable parameters---one for the interaction range and one for the interaction complexity.
\end{abstract}

\end{frontmatter}


\section{Introduction}

Accurate computational prediction of properties of alloys from their composition is one of the outstanding problems of rational materials design.
Emerging applications, such as the high-entropy alloys (HEAs) \cite{Yeh2004HEA}, pose new challenges to computational materials methods.
HEAs are defined as alloys with five or more constituent elements in equal or close to equal proportions \cite{Yeh2004HEA}.
A large number of elements leads to high configuration entropy of the solid solution phase and hence stabilizes it.
Owing to this, HEAs possess many unique mechanical properties \cite{HEA1,HEA2,HEA3}.
Accurate computational prediction of the mixing enthalpy and configuration entropy would be very instrumental in studying HEAs, as it is hard to experimentally explore different compositions of five or more elements due to combinatorial complexity.
The state-of-the-art methodology of computationally assessing the stability of multicomponent crystalline alloys is based on cluster expansion \cite{2013CE-CS,2016CE-review,Sanchez2017-CE-foundations,Nguyen2017relaxation-error}, allowing to fit formation energies of binary systems 
over the entire range of compositions, ternary and quaternary systems \cite{Ternary1,Ternary2,Ternary3,Quaternary2016} 
over, typically, some subrange of the composition range, and quinary systems at specific points of the composition range \cite{Quinary2015}.

Cluster expansion belongs to a class of data-driven interatomic interaction models, along with machine learning interatomic potentials \cite{Behler2016review,GAP2014,Shapeev2016-MTP,Thompson2015316}.
Data-driven models assume a flexible functional form of the interaction energy with many (hundred or more) free parameters that are fitted from the quantum-mechanical data.
The major drawback of cluster expansion is the ``relaxation error'' \cite{Nguyen2017relaxation-error} which refers to low ability of cluster expansion to account for changes in mixing enthalpy related to the displacements of atoms to their equilibrium positions due to, for instance, the mismatch in their atomic radii.
This is probably the reason why cluster expansion is not very successful in handling systems with a large number of components.

In this paper I propose a new approach to accurately representing and fitting the formation energies of alloys with a large number of elements.
The new approach does not directly resolve the ``relaxation error'' issue, but is shown to have a comparable or better performance than cluster expansion in cases when the ``relaxation error'' is low. 
The approach is based on partitioning the energy into contributions of the atomic environments and representing these contributions with low-rank multidimensional tensors \cite{Oseledets2011TT}.
The proposed model has only two adjustable parameters, the range of the interatomic interaction and the upper bound on the tensor rank, the latter controls the number of free parameters in the model.
The idea of a low-rank representation of functions of the atomic environments was also pursued in \cite{Zunger2011lowrank}.



\section{Interatomic Interaction}

I consider the following model of a crystal.
Let the undeformed positions of atoms be $\mA \bbZ^3$, where $\bbZ^3$ is the lattice of all points with integer coordinates and the matrix $\mA$ sets the actual crystal structure and dimensions.
For example, a face-centered cubic (f.c.c.)\ lattice with constant $a$ is defined by $\mA=\left(\begin{smallmatrix}0 & a/2 & a/2 \\ a/2 & 0 & a/2 \\ a/2 & a/2 & 0\end{smallmatrix}\right)$.
Each atom $\xi\in \mA \bbZ^3$ is displaced by $x(\xi)\in\bbR^3$ from its reference position $\xi$ and is of the type $\sigma(\xi)\in\{1,\ldots,m\}$ \footnote{in the cluster expansion literature the degrees of freedom are typically denoted by $\sigma(\xi)$, following the notation of the electron spin.}.
Let $\Omega\subset \mA \bbZ^3$ be a computational domain (supercell) repeated periodically in the entire space.
Then the degrees of freedom of the atomistic system are $x=x(\xi)$ and $\sigma=\sigma(\xi)$ for each $\xi\in\Omega$ and the interaction energy is $\tilde{E}=\tilde{E}(x,\sigma)$.
I will refer to this as an \emph{on-lattice} model of a crystal, since the atoms are indexed by their undeformed position in an ideal lattice.
This is in contrast with, for example, interatomic potentials, such as the embedded atom method (EAM) or the machine learning based ones \cite{Behler2016review,GAP2014,Shapeev2016-MTP,Thompson2015316}, that do not relate atoms to their undeformed positions.

Unless we want to model defects, the displacements $x$ can be eliminated from the model.
Often, it is assumed that the atoms are at (or near) their relaxed positions and hence
\begin{equation} \label{eq:Edef}
E(\sigma) := \min_x \tilde{E}(x,\sigma).
\end{equation}
This model is good for sampling different distributions of atoms to the lattice sites and hence, e.g., estimating the configuration entropy of an alloy with a known lattice (for which the atomic displacements are not dramatic).

The atoms are assumed to interact only with their closest environment characterized by the cut-off distance $R_\cut$ and the interaction neighborhood comprised of all vectors $r_1,\ldots,r_n\in\mA \bbZ^3$ whose length is less than $R_\cut$.
Hence, the energy of interaction of these atoms is postulated to be
\begin{equation}\label{eq:E}
E(\sigma) = \sum_{\xi\in\Omega} V(\sigma(\xi+r_1),\ldots,\sigma(\xi+r_n)),
\end{equation}
where $V$ is called, by analogy with the off-lattice case, the ``interatomic potential''.
Essentially, $V$ is an $m\times m\times\ldots\times m$, $n$-dimensional tensor that defines the interaction model \eqref{eq:E}.
The goal is to fit this model to the ``true'' quantum-mechanical model given by $E^\qm(\sigma)$.
Without loss of generality it can be assumed that $E^\qm(\sigma)=0$ whenever all $\sigma(\xi)$ are equal---in this case $E^\qm(\sigma)$ is called the formation energy of $\sigma$.


The fitting is done on a set of $K$ atomistic configurations $\sigma^{(k)}$, $k=1,\ldots,K$, given together with their true energies $E^\qm\big(\sigma^{(k)}\big)$.
Then, the sought $V$ is obtained by minimizing the mean-square functional
\begin{equation}\label{eq:pre-MS-func}
\frac{1}{K} \sum_{k=1}^K \Big|E\big(\sigma^{(k)}\big) - E^\qm\big(\sigma^{(k)}\big)\Big|^2.
\end{equation}
This is a linear regression problem on the multidimensional tensor $V$.

The typical datasets quoted in the literature have a few hundreds of configurations, whereas, for example, for an f.c.c.\ crystal even for two species ($m=2$) and 12 nearest neighbors ($n=13$) the problem \eqref{eq:pre-MS-func} has $m^n=8192$ unknown parameters fitting which already seems intractable.
Accounting, however, for physical symmetries (i.e., enforcing $V$ to be invariant with respect to the f.c.c.\ crystal symmetry space group) reduces the number of unknowns to $288$ and the problem becomes tractable.
However, if the number of species increases to $m \geq 3$ then the number of unknowns becomes of the order of $10^5$ or more, which necessitates further reduction in the number of unknowns.
Below I show that the so-called low-rank tensor representation for $V$ successfully reduces the number of unknowns in the regression problem and yields an efficient way of accurately fitting the formation energies.

\section{Formation Energy Representation}

\subsection{Low-rank Tensors}
I first review the concept of low-rank matrices and tensors.
Consider an $m\times m$ matrix $M=M(i,j)$, where $i,j\in\{1,\ldots,m\}$.
The matrix has rank $r$ or less if it can be represented as 
\[
M(i,j) = \sum_{\ell=1}^{r} u_\ell(i) \, v_\ell(j),
\]
where $u$ and $v$ are the scaled singular vectors of $M$.
As a natural generalization, we can say that a tensor $V$ has rank $r$ if 
\begin{equation}\label{eq:low_rank_simple}
V(\sigma_1, \ldots, \sigma_n) = \sum_{\ell=1}^{r} u^{(1)}_\ell(\sigma_1) \ldots u^{(n)}_\ell(\sigma_n)
\end{equation}
for some $u^{(1)}_\ell, \ldots, u^{(n)}_\ell$.
It is known that the set of tensors of the form \eqref{eq:low_rank_simple} is not a closed set, i.e., a limit of a sequence of low-rank tensors may be a high-rank tensor.
This can be illustrated, for example, by taking a simple one-body energy $V(\sigma_1, \ldots, \sigma_n) = \varphi(\sigma_1)+\ldots+\varphi(\sigma_n)$ which is a rank-$n$ tensor, but it can be approximated with an arbitrary accuracy by the following rank-two tensor:
\[
\frac{(1+\eps \varphi(\sigma_1))\ldots(1+\eps \varphi(\sigma_n))-1}{\eps}
\approx
\big(\varphi(\sigma_1)+\ldots+\varphi(\sigma_n)\big).
\]
Due to this, the low-rank tensors might not behave well when iteratively solving an optimization problem.

I will therefore use an alternative version of low-rank tensors that is free from this problem.
Namely, I will use the so-called tensor train (TT) representation \cite{Oseledets2011TT} defined by
\begin{equation}\label{eq:TT}
V(\sigma_1, \ldots, \sigma_n)
= 
A^{(1)}(\sigma_1) \ldots A^{(n)}(\sigma_n),
\end{equation}
where each $A^{(i)} = A^{(i)}(\sigma_i)$ is an $\sigma_i$-dependent matrix of size $r_{i-1}\times r_i$, and $r_0=r_n=1$.
The rank of this representation is defined as $\bar{r} := \max_i r_i.$
I will call the model \eqref{eq:TT} the low-rank potential (or LRP).
The LRP has about $n m \bar{r}^2$ free parameters, which is much less than $m^n$ as was before using the low-rank assumption.

\subsection{Fitting the Interatomic Potential}

I thus assume that the interatomic potential $V$ has the form \eqref{eq:TT}, where, after $R_\cut$ is fixed, $\bar{r}$ is the only parameter controlling the accuracy of the model.
The $\sigma$-dependent matrices $A^{(i)}$ are the unknown parameters of the model that are found by fitting to the ab initio data.
Totally, there is $O(n m \bar{r}^2)$ parameters.
It should be noted that \eqref{eq:TT} is not invariant with respect to the space group of the cubic lattice, $G$, consisting of 48 linear space transformations $g\in\bbR^{3\times 3}$ that map $\mA \bbZ^3$ into itself.
The mean-square functional \eqref{eq:pre-MS-func} is therefore adjusted to
\begin{equation}\label{eq:MS-func}
J := \frac{1}{48 K} \sum_{g\in G} \sum_{k=1}^K \Big|E\big(g\big(\sigma^{(k)}\big)\big) - E^\qm\big(\sigma^{(k)}\big)\Big|^2.
\end{equation}
The training (fitting) problem is hence:
\[
\text{
find $A^{(1)},\ldots,A^{(n)}$ minimizing $J$ subject to \eqref{eq:E} and \eqref{eq:TT}.}\vphantom{\sum}
\]

To solve this optimization problem, it should be noted that the functional $J$ is quadratic in $E$, the latter is linear in $V$, and $V$ is linear in each $A^{(i)}$. 
I hence use the alternating least squares (ALS) algorithm (see, e.g., \cite{TT-ALS}), consisting of taking an initial guess for all $A^{(i)}$ and then updating $A^{(i)}$ in an iterative manner until convergence.
Each iteration consists of $n$ subiterations, where in the $i$-th subiteration $J$ is minimized with respect to $A^{(i)}$ while freezing $A^{(j)}$, $j\ne i$.
The latter is a standard quadratic optimization problem which is equivalent to solving a system of linear algebraic equations and is solved by the Gauss--Seidel method.
As a final component of the algorithm, the simulated annealing is used in order to avoid the iterations getting stuck in local minima.
%

The iterations are stopped when the difference in $J$ between two consecutive iterations is less than a certain threshold.
The value $\sqrt{J}$ then is the training root-mean-square error (RMSE).
In case if the training set size, $K$, is small, it can be significantly smaller than the actual prediction (validation) error of the model.
To assess the latter, a validation set $\tilde{x}^{(1)},\ldots,\tilde{x}^{(\tilde{K})}$ is generated independently of the training set and the validation RMSE is then $\tilde{J}^{1/2}$, where
\[
\tilde{J} := \frac{1}{48 \tilde{K}} \sum_{g\in G} \sum_{k=1}^{\tilde{K}} \Big|E\big(g\big(\tilde{x}^{(k)}\big)\big) - E^\qm\big(\tilde{x}^{(k)}\big)\Big|^2.
\]
In what follows, by the fitting error I mean the validation RMSE.


\section{Numerical Tests}

To test the proposed model for representing the formation energy, LRP, I calculate and fit the formation energies of a number of systems using the density functional theory (DFT) as implemented in the VASP package \cite{VASP1,VASP4}, the projected augmented wave (PAW) pseudopotentials \cite{Blochl1994PAW}, and the Perdew-Burke-Ernzerhof exchange-correlation functional \cite{PBE}.
In some tests the atomistic configurations were relaxed: the system was driven to a local minimum with respect to the atomic positions and the supercell size and shape. 
Unless stated otherwise, the atomistic configurations were generated as follows.
The computational cell was a cube with the side of 8\AA$\mathstrut$ containing 32 atoms arranged in an f.c.c.\ lattice with the lattice constant of 4\AA$\mathstrut$.
The $\Gamma$-centered $4\!\times\!4\!\times\!4$ k-point mesh was used.
This ensures that the DFT energies are converged to about 1.3 meV/atom accuracy.
The configurations with $m$ species were generated randomly and independently from each other, by first uniformly sampling $m$ numbers, $n_1,\ldots,n_m$ such that $n_1+\ldots+n_m = 32$, and then for each species $i$, $n_i$ atoms of that species are placed in the free sites of the lattice.
This ensures that the entire range of alloy compositions is properly covered.

For each test the two sets of configurations were generated: the training set of varying size which was used for the fitting and the validation set of 200 to 400 configurations that were not involved in the fitting and on which the prediction error was measured.
In what follows, ``error'' and ``accuracy'' refers to this prediction error.
The fitting was then done with a sequence of $\bar{r}=2,3,\ldots$, and the value of $\bar{r}$ for which the 10-fold cross-validation error \cite{Alpaydin2014MachineLearning} was minimal was chosen.
Cross-validation is a technique of estimating the validation error based only on the training set.
Unless stated otherwise, the 13-atom nearest-neighbor model was used.

\begin{figure}[htb]
\begin{center}
\includegraphics[scale=.78]{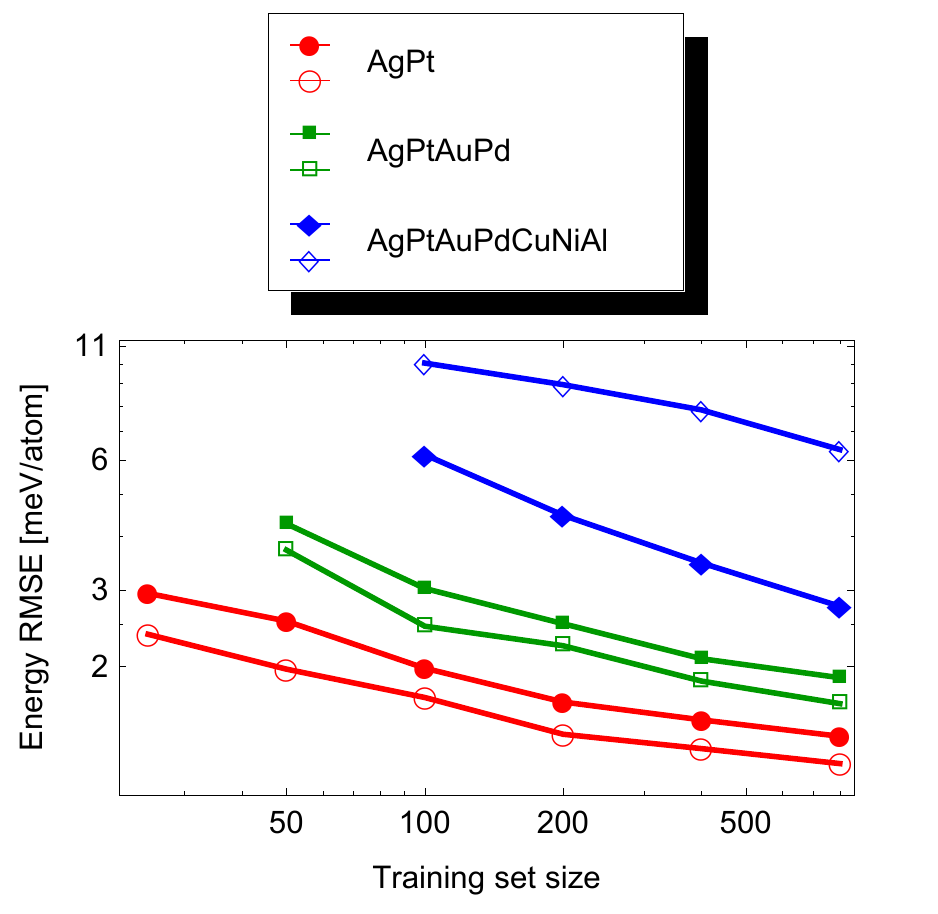}
\end{center}
\caption{%
The learning curves (i.e., the root-mean-square energy error as measured on the validation set as a function of the training set size) for three different systems.
The filled markers correspond to the fitting of unrelaxed configurations, while the hollow markers correspond to the fitting of relaxed ones.
}
\label{fig:big_conv}
\end{figure}

\subsection{Convergence tests}

The first test demonstrates an excellent performance of LRP on a sequence of three systems, Ag-Pt, Ag-Pt-Au-Pd, and Ag-Pt-Au-Pd-Cu-Ni-Al.
The results of this test are shown in Fig.\ \ref{fig:big_conv}.
When fitting the formation energy of the unrelaxed configurations, the training set size of approximately 25, 100, or 600 configurations was sufficient to reach the accuracy of 3 meV/atom for, respectively, the binary, quaternary, and septenary systems.
When fitting the energy of the relaxed Ag-Pt or Ag-Pt-Au-Pd configurations, the error was even smaller than that for the unrelaxed configurations.
In contrast, the fitting error of the relaxed septenary alloy configurations was about twice larger than that of the relaxed configurations.
The latter may be attributed to large relaxations in the Ag-Pt-Au-Pd-Cu-Ni-Al system due to significantly different sizes of atoms, which a displacement-free model cannot capture.

\begin{table}
\begin{center}
\begin{tabular}{|l|c|c|}\hline
system & training error & prediction error \\\hline
AuPd (unrelaxed) & 1.58 {\small(1.8\%)} & 1.71 {\small(1.9\%)} \\
AuPt (unrelaxed) & 1.39 {\small(11.8\%)} & 1.53 {\small(12.9\%)} \\
AgPd (unrelaxed) & 1.42 {\small(2.6\%)} & 1.60 {\small(2.9\%)} \\
AgPt (unrelaxed) & 1.34 {\small(9.6\%)} & 1.38 {\small(9.3\%)} \\
AgPt (relaxed) & 1.19 {\small(3.0\%)} & 1.19 {\small(3.0\%)} \\
AgPtAuPd (unrelaxed) & 1.47 {\small(2.4\%)} & 1.89 {\small(3.1\%)}  \\
AgPtAuPd (relaxed) & 1.25 {\small(1.5\%)} & 1.64 {\small(1.9\%)}  \\
AgPtAuPdCuNiAl (unrelaxed) & 1.55 {\small(0.7\%)} & 2.76 {\small(1.2\%)} \\
AgPtAuPdCuNiAl (relaxed) & 3.10 {\small(0.8\%)} & 6.34 {\small(1.6\%)} \\
\hline
\end{tabular}
\caption{Training error and prediction errors for the different systems when fitted on 800 training configurations.
The absolute errors are quoted in meV/atom and the relative errors (in parenthesis) and in \%.}
\label{tab:errors}
\end{center}
\end{table}

The absolute and relative quantities of training and prediction errors for different systems are given in Table \ref{tab:errors}.
It can be seen that the absolute error has a much lower difference for different systems, as compared to the relative error.
Interestingly, the ``harder'', septenary system has a very small relative error.

The following tests are intended to study the approximation properties of LRP and therefore were all done on the unrelaxed configurations.

\begin{figure}[htb]
\begin{center}
\includegraphics[scale=.78]{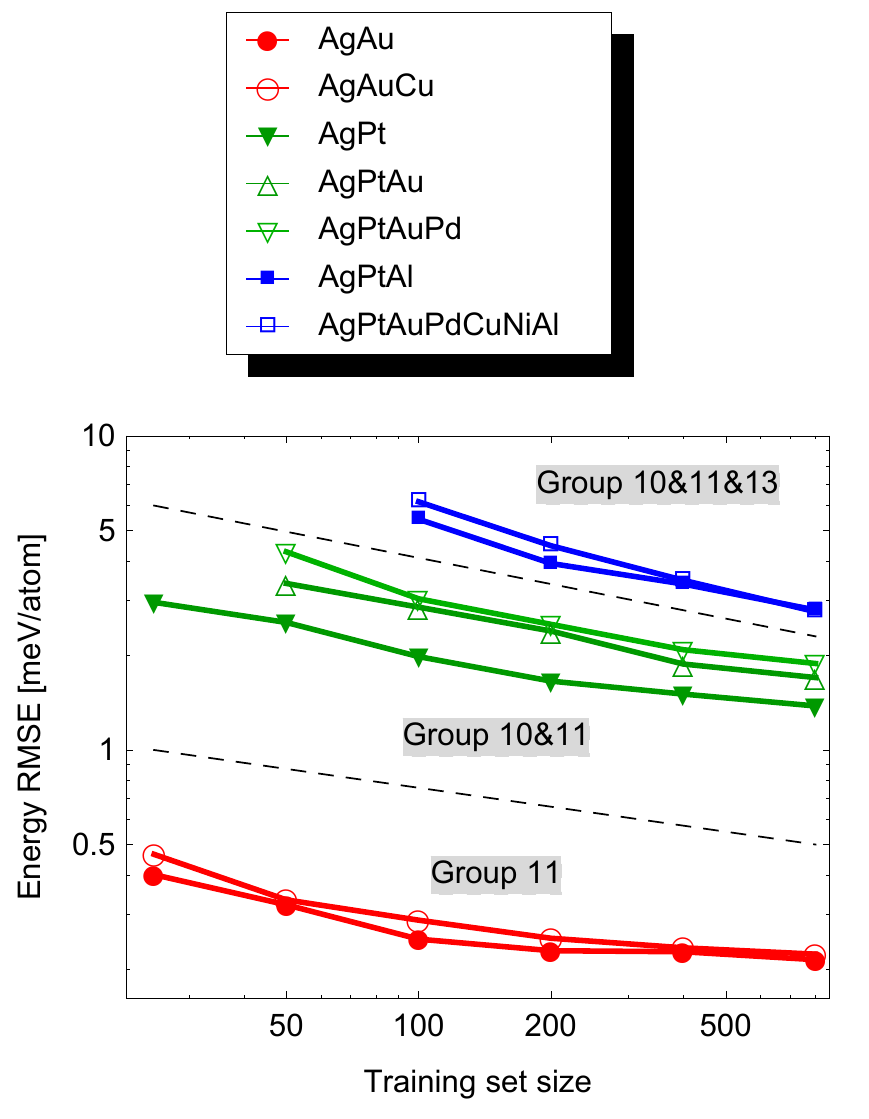}
\end{center}
\caption{%
The learning curves for seven different systems.
The error depends strongly on the number of groups that the elements belong to.
}
\label{fig:columns}
\end{figure}

\subsection{Mixing elements from different columns}

In the next test we will see that the fitting error shows stronger dependence on the number of groups (i.e., periodic table columns) that the alloy elements belong to, and weaker dependence on the number of elements themselves.
As can be seen from Fig.\ \ref{fig:columns}, the fitting error of the ternary system of group 11 elements is smaller than that of the Ag-Pt system, and the error of the quaternary system with groups 10 and 11 elements is smaller than that of the Ag-Pt-Al system.
The latter error is essentially the same as for the septenary system with elements from the three groups.
Another remarkable property that can be seen in Fig.\ \ref{fig:columns} is that the errors for the systems with elements from the same group is an order of magnitude smaller than that for the other systems.


\begin{figure}[htb]
\begin{center}
\includegraphics[scale=.78]{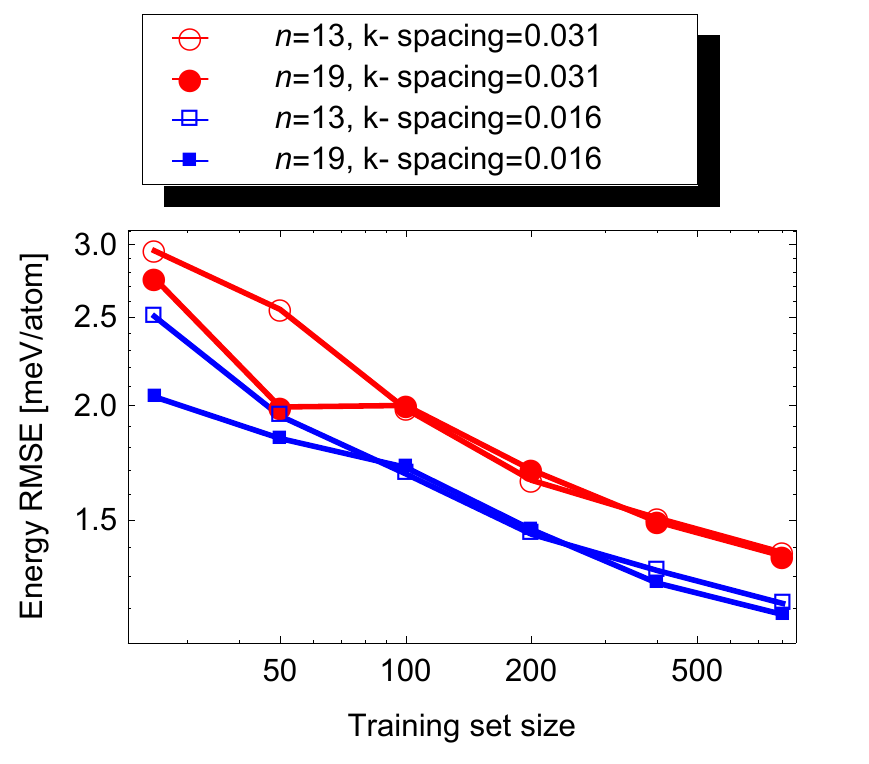}
\end{center}
\caption{%
Learning curves for different interaction ranges ($n=13$ and $n=19$) and different DFT accuracies (k-point spacing of $0.031{\rm \AA}^{-1}$ and $0.016{\rm \AA}^{-1}$).
The graphs indicate that the error cannot be reduced further without increasing the interaction range.
}
\label{fig:acc}
\end{figure}

\subsection{Is fitting with sub-meV accuracy possible?}
The results shown in Fig.\ \ref{fig:columns} naturally raise the following question: can the systems with elements from two or more groups be fitted with the accuracy lower than 1 meV/atom by refining the k-points mesh and possibly modifying the fitting scheme?
The results below suggest that it is not possible without vastly increasing $R_\cut$ (and therefore the training set size).

Four different fits were obtained for the Ag-Pt system: two models, the one with $n=13$ (first-nearest-neighbor model) and the other one with $n=19$ (next-nearest-neighbor model), were fitted to two sets of QM data, the one with $4\times 4\times 4$ k-point mesh (k-point spacing of $0.031{\rm \AA}^{-1}$) and the other with $8\times 8\times 8$ k-point mesh (k-point spacing of $0.016{\rm \AA}^{-1}$). The latter k-point mesh ensures that the DFT energies are converged to the accuracy of about $0.1$ meV/atom.
The corresponding learning curves (i.e., the error as a function of the training set size $K$) are shown in Fig.\ \ref{fig:acc}.
These results indicate that including the next nearest neighbors does not significantly improve the results, and fitting to highly converged DFT energies improves the error only slightly, by about 10\%.

Finally, the nearest-neighbor model fitted without the low-rank restriction on 800 Ag-Pt configurations was compared to the low-rank fit.
Without the low-rank restriction the model has 288 coefficients that were fitted to 800 energies.
The validation error of this model was $2.15$ meV/atom and the fitting error was $1.0$ meV/atom.
This indicates that lifting the low-rank restriction cannot significantly improve the error.


The results above confirm that a large part of the remaining 1--3 meV/atom errors shown in Fig.\ \ref{fig:big_conv} is due to nonlocality in the interatomic interaction.
This nonlocality can be explained by the fact that the electron density perturbation in metals decays slowly \cite{Fridel-screening}.
This means that replacing an atom in a configuration by another atom with a different number of valence electrons creates a nonlocal perturbation, whereas replacing atoms while keeping the same number of valence electrons introduces a more local perturbation.
This is in agreement with the fact that the Ag-Au and Ag-Au-Cu systems could be fitted much more accurately compared to the other systems.

\subsection{Comparison with Cluster Expansion}

\begin{figure}[htb]
\hfill
\includegraphics[width=20em]{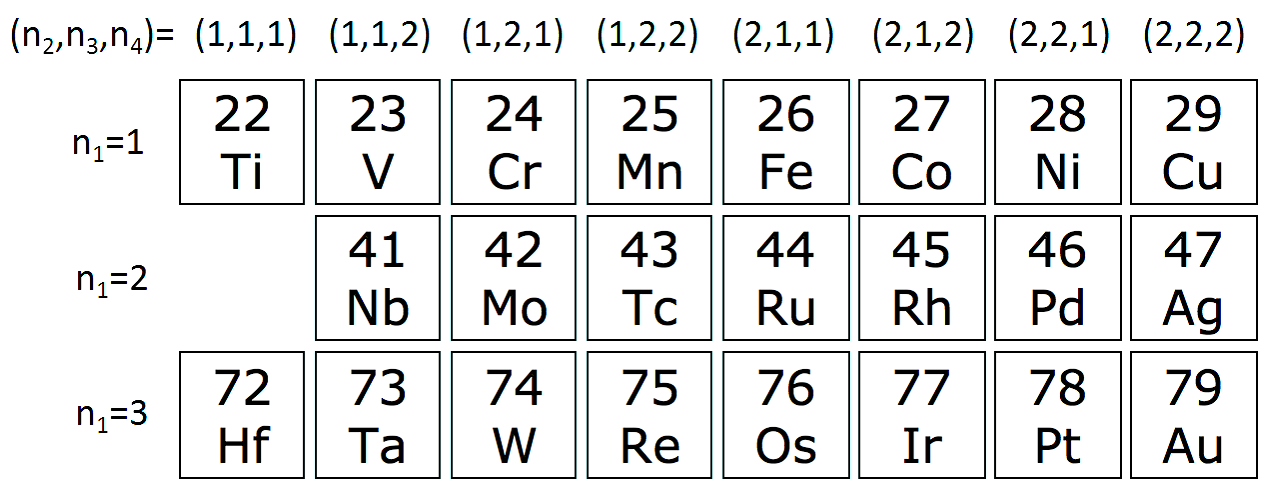}
\hfill
$\mathstrut$
\caption{%
The 23 elements, as they appear in the periodic table, for mixtures of which the formation energy was fitted.
Each element is indexed with four integer numbers, $(n_1,n_2,n_3,n_4)$.
}
\label{fig:23-elements}
\end{figure}

Finally, to test how the performance of LRP scales with the number of components, I fit the formation energy of alloys with up to 23 elements, listed in Fig.\ \ref{fig:23-elements}, with LRP and with the cluster expansion.
The element Zr (atomic number 40) was excluded in order to eliminate a possible source of errors due to the choice of pseudopotentials---a pseudopotential with 10 electrons was chosen for Ti and Hf, but a similar pseudopotential for Zr was not available. 
A training set with 1600 configurations and a validation set with 200 configurations was generated.

For the cluster expansion model, all two-body and three-body clusters of nearest neighbors were chosen, totaling to 2576 unknown parameters.
Since this number is larger than the largest training set size, I have used the compressive sensing cluster expansion as proposed in \cite{2013CE-CS}.
The LRP model is slightly modified for this test case: instead of indexing the types of atoms with numbers from 1 to 23, they are instead indexed with four integer numbers $(n_1, n_2, n_3, n_4)$, as shown in Fig.\ \ref{fig:23-elements}.
By choosing this way of indexing the atomic types, I implicitly assume some kind of regularity of dependence of the interaction energy on the position of the element in a periodic table.
The interatomic potential $V$ is hence a $13\times 4 = 52$--dimensional tensor of size 2 or 3 in each dimension.

\begin{figure}[htb]
\hfill
\includegraphics[width=20em]{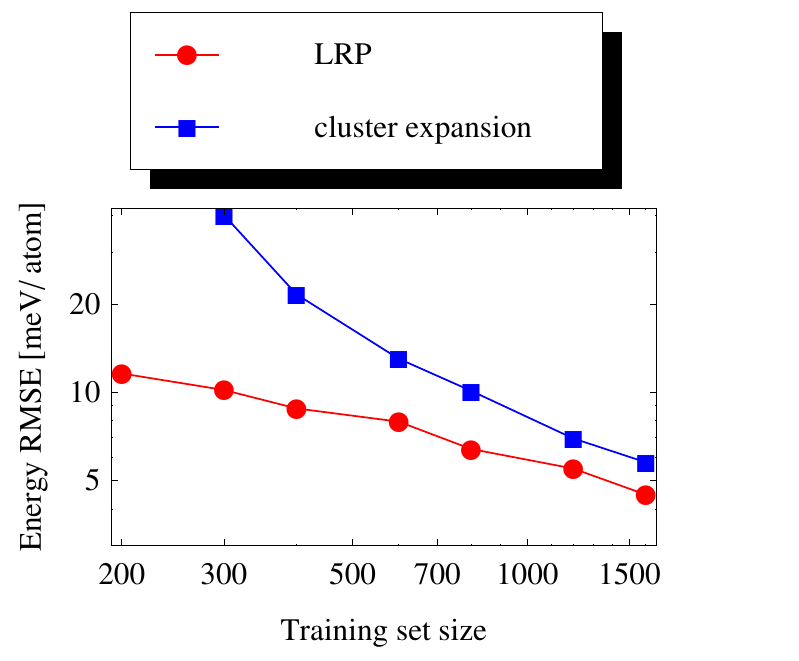}
\hfill
$\mathstrut$
\caption{%
Learning curves for fitting formation energies of alloys with up to 23 elements.
}
\label{fig:learning-curve-23}
\end{figure}

The learning curves are shown in Fig.\ \ref{fig:learning-curve-23}.
It can be seen that LRP shows significantly lower errors than cluster expansion for relatively small training set sizes, and the two algorithms start showing comparable errors when the training set size becomes larger than about 1000.



\section{Discussion and Conclusion}

The proposed low-rank potential (LRP) demonstrates a comparable or better performance in the fitting of multicomponent alloys than cluster expansion in the case when the errors due to relaxation are small.
The proposed model is simple and has only two adjustable parameters---one to control the range of the interaction and the other one to control the number of fitting parameters.
The fitting procedure needs nonlinear optimization, however, the evaluation of the fitted model is very fast: it reduces to performing, for each atom, a few multiplications of a small matrix by a vector.
When the number of components is five or less, the entire tensor $V$ encoding the interaction can be tabulated and make the model evaluation extremely fast.

Although, as was shown in this paper, LRP can approximate the interatomic interaction very accurately and efficiently, the error due to relaxation of atoms may still be large.
Indeed, Fig.\ \ref{fig:big_conv} shows that when the atomic sizes are significantly different, the accuracy of fitting the energy of the relaxed configurations is significantly less than that of the unrelaxed configurations---this is similar to how cluster expansion behaves \cite{Nguyen2017relaxation-error}.
This can be rectified in part by introducing, for instance, the explicit dependence of the energy on the average composition-dependent equilibrium volume of the lattice \cite{Sanchez2017-CE-foundations}.
Moreover, when some elements have a different ground state lattice (e.g., body-centered cubic), the model \eqref{eq:Edef} may be unusable for certain compositions.
It can be hypothesized that in this case there may be several local minima that the atomic configurations may be relaxed to (due to symmetry breaking of the f.c.c.\ lattice), which would imply that the energy of the ground state is not a function of $\sigma$.
In this case, including $x$ back in the model may rectify its behavior.

\subsection*{Acknowledgments}
The author thanks Prof.\ J\"org Neugebauer, Prof.\ G\'abor Cs\'anyi and Prof.\ Alexandre Tkatchenko for valuable discussions.
This work was supported by the Skoltech NGP Program No.\ 2016-7/NGP (a Skoltech-MIT joint project).
A part of the work was done by the author during the Fall 2016 long program at the Institute of Pure and Applied Mathematics, UCLA.

\bibliography{paper}

\end{document}